\documentclass{aa}
\usepackage{txfonts}
\usepackage{graphicx}
\usepackage{multicol}
\usepackage{multirow}
\begin{document}

   \title{The contact binary VW Cephei revisited: surface activity and period variation}
   
   \authorrunning{T. Mitnyan, A. Bódi, T. Szalai et al.}
   \titlerunning{The contact binary VW Cephei revisited}


   \author{T. Mitnyan\inst{1}, A. Bódi\inst{2,3}, T. Szalai\inst{1}, J. Vinkó\inst{1,3}, K. Szatmáry\inst{2}, T. Borkovits\inst{3,4}, B. I. Bíró\inst{4}, T. Hegedüs\inst{4}, K. Vida\inst{3}, A. Pál\inst{3}
          }

   \institute{Department of Optics and Quantum Electronics, University of Szeged, H-6720 Szeged, Dóm tér 9, Hungary
         \and
         Department of Experimental Physics, University of Szeged, H-6720 Szeged, 
Dóm tér 9, Hungary
\and
             Konkoly Observatory, Research Centre for Astronomy and Earth Sciences, Hungarian Academy of Sciences, H-1121 Budapest, Konkoly Thege Mikl\'os \'ut 15-17, Hungary
\and 
Baja Astronomical Observatory of University of Szeged, H-6500 Baja, Szegedi út, Kt. 766, Hungary\\
             \email{mtibor@titan.physx.u-szeged.hu}
             }
             
   \date{Received ...; accepted ...}

 
  \abstract
   {Despite the fact that VW Cephei is one of the well-studied contact binaries in the literature, there is no fully consistent model available that can explain every observed property of this system.}
   {Our motivation is to obtain new spectra along with photometric measurements, to analyze what kind of changes may have happened in the system in the past two decades, and to propose new ideas for explaining them.}
   {For the period analysis we determined 10 new times of minima from our light curves, and constructed a new O$-$C diagram of the system. Radial velocities of the components were determined using the cross-correlation technique. The light curves and radial velocities were modelled simultaneously with the \texttt{PHOEBE} code. All observed spectra were compared to synthetic spectra and equivalent widths of the H$\alpha$ line were measured on their differences.}
   {We have re-determined the physical parameters of the system according to our new light curve and spectral models. We confirm that the primary component is more active than the secondary, and there is a correlation between spottedness and the chromospheric activity. We propose that flip-flop phenomenon occurring on the primary component could be a possible explanation of the observed nature of the activity. To explain the period variation of VW Cep, we test two previously suggested scenarios: presence of a fourth body in the system, and the Applegate-mechanism caused by periodic magnetic activity. We conclude that although none of these mechanisms can be ruled out entirely, the available data suggest that mass transfer with a slowly decreasing rate gives the most likely explanation for the period variation of VW Cep.}
   {}


   \maketitle

\section{Introduction}
A contact binary typically consists of two main sequence stars of the F, G or K spectral type that are orbiting around their common center of mass. According to the most accepted model, both components are filling their Roche lobes and making physical contact through the inner (L1) Lagrangian point. The components also share a common convective envelope \citep{lucy68}, which is the reason why they have nearly the same effective temperature. Although this model is very simple and it can explain a lot of properties of these stars, it leaves some open questions \citep{rucinski10}. The most obvious one is that there are a lot of systems where the mass ratio determined from photometry differs from the one obtained from spectroscopy. Nevertheless, a general model that resolves all the issues is still lacking, which may give further motivation for studying such kind of binary systems.

The light curves of contact binaries show continuous flux variation, even outside of eclipses, because of the heavy geometric distortion of both components. The orbital periods of these objects are usually shorter than a day. There are two main types of these objects according to their light curves: A-type and W-type \citep{binnendijk70}. In A-type systems, the more massive primary component has higher surface brightness than the secondary. They have longer orbital periods and smaller mass ratios than the W-type systems, where the secondary component seems to show higher surface brightness. The most accepted explanation for the existence of W-type objects is that the more massive, intrinsically hotter primary component is so heavily spotted that its averaged surface brightness appears to be lower than that of the less massive and cooler secondary \citep{mullan75}. Nevertheless, another scenario, the hot secondary model, also commonly used in the literature in order to explain the light curves of these binaries; in this model, it is assumed that the secondary star has a higher temperature for an unknown reason \citep{rucinski74}.

A lot of contact binaries show asymmetry in the light curve maxima (O'Connell effect), which is likely caused by starspots on the surface of the components as a manifestation of the magnetic activity. The difference between the maxima can change from orbit to orbit because of the motion and evolution of these cooler active regions. This phenomenon may indicate the presence of an activity cycle similarly to what can be seen on our Sun.

Activity cycles have been directly observed on several types of single and binary stars \citep[e.g.][]{wilson78,baliunas95, heckert95,olah02,vida15}. In the cases of contact binaries there is only indirect evidence for periodic magnetic activity. These mostly come from period analysis, which have revealed that the periodic variation between the observed and calculated moments of mid-eclipses cannot be simply explained with the presence of additional components in the system (or, it can only be a partial explanation). \citet{applegate92} proposed a mechanism that could be responsible for such kind of cyclic period variations. According to his model, the magnetic activity can modify the gravitational quadrupole moment of the components that induces internal structural variations in the stars, which also influences their orbit.

Many contact binaries also have a detectable tertiary component. In fact, \citet{pribulla06} summarized that up to V=10 mag, $59\pm8\%$ of these systems are triples on the northern sky, which becomes $42\pm5\%$, if the objects on the southern sky are also included. These numbers are based on only the verified detections, but they can increase up to $72\pm9\%$ and $56\pm6\%$, respectively, if one takes into account even the non-verified cases. These numbers are just lower limits because of unobserved objects and the undetectable additional components; therefore, the authors suggested that it is possible that every contact binary is a member of a multiple system.  According to some models \citep[e.g.][]{eggleton01}, the tertiary components can play an important role in the formation and evolution of these systems.


VW Cephei, one of the most frequently observed contact binary, was discovered by \citet{schilt26}. It is a popular target
because of its brightness (V = 7.30 - 7.84 mag) and its short period ($\sim$ 6.5 hours), which allows to easily observe a full orbital cycle on a single night with a relatively small telescope. The system is worth observing, because it usually shows O'Connell effect and clearly has a third component confirmed by astrometry \citep{hershey75}; however, there is no perfect model that can explain all the observed properties of the system.

The strange behaviour of the light curve of VW Cephei was discovered during the first dedicated photometric monitoring of the system \citep{kwee66}. Several different models have been developed to explain the asymmetry of maxima e.g. the presence of a circumstellar ring \citep{kwee66}, a hot spot on the surface caused by gas flows \citep{vantveer73}, or cool starspots on the surface \citep{yamasaki82}. The latter model seemed to be the most consistent with the observations and dozens of publications presented a light curve model including starspots.

Instead of the huge amount of photometric data, there is a limited number of published spectroscopic observations about VW Cep. The first spectroscopic results are from \citet{popper48} who published a mass ratio of $q$ = 0.326 $\pm$ 0.045. \citet{anderson80} took two spectra at two different orbital phases and derived a mass ratio of 0.40 $\pm$ 0.05, which was consistent with the previous value of 0.409 $\pm$ 0.011 yielded by \citet{binnendijk67}. \citet{hill89} measured the radial velocities of the components using the cross-correlation technique on low-resolution spectra, and determined a 
new value of $q$ = 0.277 $\pm$ 0.007. The next spectroscopic study of the system was presented by \citet{frasca96} who took a series of low resolution spectra at the H$\alpha$-line; they showed that there is a rotational modulation in the equivalent width of the H$\alpha$-profile. \citet{kaszas98} obtained the first series of medium resolution spectra of the system. They used the same technique as \citet{hill89} to derive the radial velocities and the mass ratio, but they got a definitely larger value ($q$ = 0.35 $\pm$ 0.01). They also investigated the variation of H$\alpha$ equivalent widths with the orbital phase and found that the more massive primary component showed emission excess, which they interpreted as a consequence of enhanced 
chromospheric activity. They pointed out a possible anticorrelation between the photospheric and chromospheric activity. The latest publication about VW Cephei including spectroscopic analysis is a Doppler imaging study by \citet{hendry00}. They derived a mass ratio of $q$ = 0.395 $\pm$ 0.016 and found that the surfaces of both components were heavily spotted, and the more massive primary component had an off-centered polar spot at the time of their observations.


No other papers based on optical spectroscopy were published about VW Cephei after the millenium. However, there were two publications based on X-ray spectroscopy \citep{gondoin04,huenemoerder06}, and another one based on UV spectra \citep{sanad14}. \citet{gondoin04} concluded that VW Cephei has an extended corona encompassing the two components and it shows flaring activity. \citet{huenemoerder06} also detected the flaring activity of the system and that the corona is mainly on the polar region of the primary component. \citet{sanad14} studied UV emission lines and found both short- and long-term variations in the strength of these lines, which they attributed to the chromospheric activity of the primary component.

In this paper, we perform a detailed optical photometric and spectroscopic analysis of VW Cep, and examine what kind of changes may have occured in the past two decades. The paper is organized as follows: Section 2 contains information about the observations and data reduction process, as well as the steps of the analysis of our photometric and spectroscopic data. In Section 3, we discuss the activity of VW Cep based on our modeling results, then, in Section 4, we present an O$-$C analysis of the system. Finally, we briefly summarize our work and its implications in Section 5.

\section{Observations and analysis}

\subsection{Photometry}

\hspace{0.5cm}Photometric observations took place on 3 nights in August 2014, and 2 nights in April 2016 at Baja Astronomical Observatory of University of Szeged (Hungary) with a 0.5m, f/8.4 RC telescope through SDSS {\it g'r'i'} filters. In order to reduce photometric errors, bias, dark and flat correction images were taken on every observing night. The image processing was performed using \texttt{IRAF}\footnote{Image Reduction and Analysis Facility: http://iraf.noao.edu}: we carried out image corrections and aperture photometry using the \texttt{ccdred} and \texttt{apphot} packages, respectively. The photometric errors are determined by the \texttt{phot} task of \texttt{IRAF}. These errors are in the order of 0.01-0.02 mag, hence they are smaller than the symbols we used in our plots. For differential photometry, we used HD 197750 as a comparison star; although, it is considered as a variable in the SIMBAD catalog, we could not find any significant changes in its light curve compared to the check star, BD+74 880 during our observations. The $\Phi$ = 0.0 phase was assigned to the secondary minimum (when the more massive primary component is eclipsed), while the secondary minimum time of the last observing night of both years was used for phase calculation for each season.

\subsection{Spectroscopy}

Spectroscopic observations were obtained on 12 and 13 April 2016 with a $R$ = 20\,000 échelle spectrograph mounted on the 1m RCC telescope at the Piszkéstető Mountain Station of Konkoly Observatory, Hungary. The integration time was set to 10 minutes as a compromise between signal-to-noise ratio and temporal resolution. ThAr spectral lamp images were taken after every third object spectrum for wavelength calibration. Three spectra of an A0V star (*81 UMa) and of a radial velocity standard (HD 114215) were taken on each night for removing telluric lines, and for radial velocity measurements, respectively. A total of 69 new spectra of VW Cephei was obtained on these two consecutive nights with an average S/N of 34. The steps of spectrum reduction and wavelength calibration were done by the corresponding \texttt{IRAF} routines in the \texttt{ccdred} and the \texttt{echelle} packages. We divided all spectra with the previously determined blaze function of the spectrograph, then we fit all apertures with a low-order Legendre polynomial to correct the small remaining differences from the continuum. The obtained spectra were corrected for telluric spectral lines in the following way. We assumed that the averaged spectrum of a rapidly rotating A0V star contains only a few rotationally broadened spectral lines and the narrow atmospheric lines. We eliminated these broad spectral lines from the spectrum of the standard star, which resulted in a continuum spectrum containing only telluric lines and we divided the object spectra with this pure telluric spectrum. The spectra of both VW Cep and the telluric standard star have a moderate S/N; therefore, after dividing the object spectra with the pure telluric spectrum, we get corrected spectra with even lower S/N values. To help on this issue and carry out a better modelling process, we performed a Gaussian smoothing on all corrected spectra of VW Cep as a final step with the \texttt{IRAF} task \texttt{gauss}, which helps filtering out artificial features caused by noise. The FWHM of the Gaussian window was chosen to be 0.7 \mbox{\AA}, which was narrow enough to avoid the smearing of the weaker, but real lines in the smoothed spectra.

\subsection{Radial velocities}

\begin{figure}
   \centering
   \includegraphics[width=9cm]{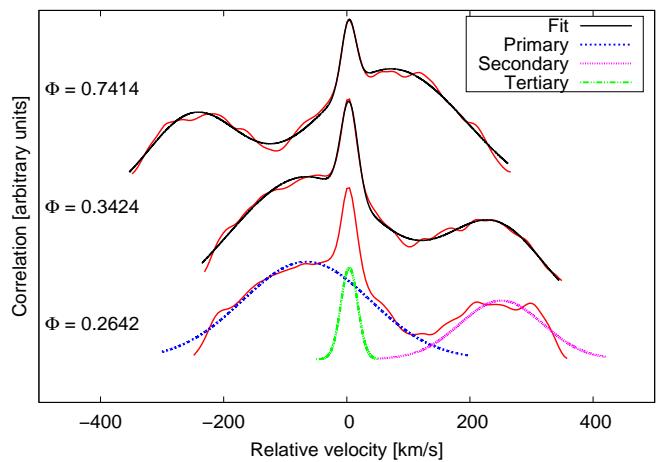}
   \caption{A sample of cross-correlation functions of different orbital phases and their fits with the sum of three Gaussian functions. We also indicated the peaks raised by different components with different lines and colours.}
   \label{ccf}
   \end{figure}

\begin{figure}
   \centering
   \includegraphics[width=9cm]{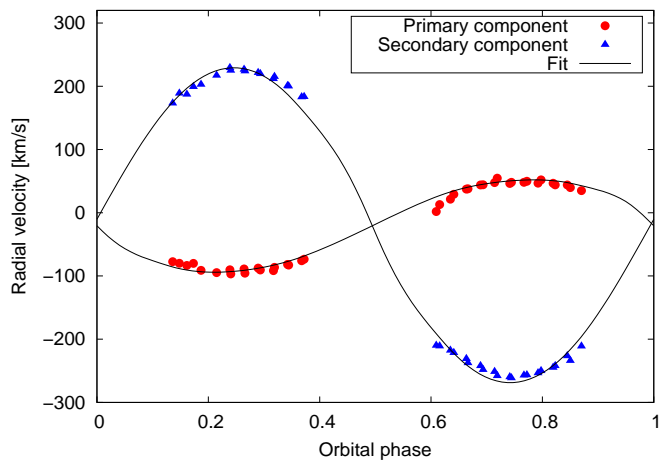}
   \caption{Radial velocity curve of VW Cephei with the fitted \texttt{PHOEBE} model. We note that the formal errors of the single velocity points (see Table \ref{radseb}) are smaller than their symbols.}
   \label{radvel}
   \end{figure}

Radial velocities of the binary components were determined using the cross-correlation technique. Each spectrum (before the smoothing process) was cross-correlated with the averaged radial velocity standard spectrum on the 6000 $-$ 6850 \mbox{\AA} wavelength region excluding the H$\alpha$-line and the region of 6250 $-$ 6350 \mbox{\AA}. This range we used is indeed narrower than our full wavelength coverage. The spectrograph approximately covers the range of 4000 $-$ 8000 \mbox{\AA}, but, because of the necessarily short exposure times and of the moderate seeing on our observation nights, the signal to noise ratio is not so high. The S/N is decreasing going toward shorter wavelengths which degrades the quality of the CCF profiles, hence we excluded the region under 6000 \mbox{\AA}. We did the same in the region between 6250 $-$ 6350 \mbox{\AA} and above 6850 \mbox{\AA}, where the strong telluric lines totally cover the features that belong to VW Cep.
Nevertheless, we note that the wavelength range we used is seven times wider than the one in \citet{kaszas98}, who also applied the CCF method to determine the radial velocities of VW Cep. The obtained CCFs were fitted with the sum of three Gaussian functions. A sample of the CCFs along with their fits are plotted in Fig. \ref{ccf}. Finally, heliocentric correction was performed on the derived radial velocities to eliminate the effects of the motion of the Earth. The radial velocities are collected in Table \ref{radseb}. The errors show the uncertainties of the maxima of the fitted Gaussians. To test the accuracy of our measurements, we derived the radial velocity of HD 114215. We got -25.35 $\pm$ 0.34 km/s, which only slightly differs from the value found in the literature \citep[-26.383 $\pm$ 0.0345 km/s,][]{soubiran13}. We also plotted them in the function of the orbital phase and fitted the velocity points of each components with the light curves simultaneously using \texttt{PHOEBE}. The radial velocity curve is plotted in Fig. \ref{radvel}.

Our newly derived mass ratio of $q$ = 0.302 $\pm$ 0.007 is inside the wide range of the previously published results, but it is slightly lower than the majority of them. Possible reasons of this disagreement are the application of different radial velocity measuring techniques, and the different quality of the data. Other authors, like \citet{popper48} and \citet{binnendijk67} measured the wavelengths of several spectral lines on low-quality spectrograms to calculate the radial velocities, while \citet{anderson80} derived the broadening functions of the components in only two different orbital phases to measure the mass ratio. \citet{hendry00} did not describe their method to derive the mass ratio, so it's difficult to compare our results to theirs. There are only two papers in the literature in which the cross-correlation technique is used to derive the radial velocities of the system. \citet{hill89} used low-resolution spectra and his CCFs show only two components, so, he probably measured his velocities on the blended profile of the primary and tertiary components, which means that his value of the mass ratio is not reliable. \citet{kaszas98} applied exactly the same method as we did; however, there are differences in the spectral range used for the cross-correlation, in the spectral resolution and in the S/N of the data: while the average S/N of their spectra is approximately 2-3 times higher than that of ours, their spectra have a spectral resolution of only $R$ $\approx$11\,000, and we used a seven times wider spectral range to calculate CCF profiles, which is a very important factor in using this method. For checking the limitations of the above described method, we applied this technique on the theoretical broadening functions of the system (see in Appendix B), which were calculated with the \texttt{WUMA4} program \citep{rucinski73}.

\subsection{Light curve modeling}

The light curves were simultaneously analyzed with the radial velocities using the Wilson-Devinney code based program \texttt{PHOEBE}\footnote{PHysics Of Eclipsing BinariEs: http://phoebe-project.org/1.0/} \citep{prsa05}, in order to create a physical model of VW 
Cephei and to determine its physical parameters. The surface-averaged effective temperature of the primary component ($T_\mathrm{eff,1}$) and the relative temperatures of the starspots were kept fixed at 5050 K as in \citet{kaszas98}, and 3500 K as in \citet{hendry00}, respectively. We assumed that the orbit is circular and the rotation of the components is synchronous, therefore, the eccentricity was fixed at 0.0, and the synchronicity parameters were kept at 1.0. For the gravity darkening coefficients and bolometric albedos, we adopted the usual values for contact systems: $g_{1}$ = $g_{2}$ = 0.32 \citep{lucy67} and $A_{1}$ = $A_{2}$ = 0.5 \citep{rucinski69}, respectively. The following parameters were varied during the modeling: the effective temperature of the secondary component ($T_\mathrm{eff,2}$), the inclination ($i$), the mass ratio ($q$), the semi-major axis ($a$), the gamma velocity ($v_{\gamma}$), the luminosities of the three components, and the coordinates and radii of starspots. The limb darkening coefficients were interpolated for every iteration from \texttt{PHOEBE}'s built-in tables using the logarithmic limb darkening law.

\begin{figure}[!h]
   \centering
   \includegraphics[width=9cm]{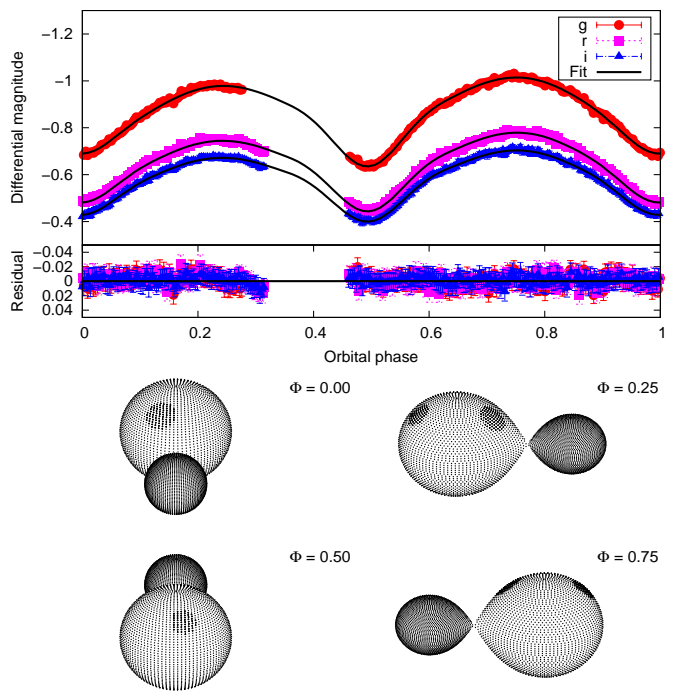}
   \caption{SDSS {\it g'r'i'} light curves of VW Cephei on 8th August 2014 along with the fitted \texttt{PHOEBE} models and the corresponding geometrical configuration and spot distribution in different orbital phases.}
   \label{20140808lc}
   \end{figure}

Light curves obtained on the given nights are plotted in Figs. \ref{20140808lc}, \ref{20140809lc}, \ref{20140810lc}, \ref{20160420lc}, and \ref{20160421lc} along with their model curves fitted by \texttt{PHOEBE} and the corresponding geometrical configurations in different orbital phases. The residuals are mostly in the range of $\pm$0.02 magnitudes, which is comparable to our photometric accuracy. One can see that the difference of the two maxima changes slightly from night to night in each filter. This variation suggests ongoing surface activity, and it is treated by assuming cooler spots on the surface of the components. The final model and spot parameters are listed in Tables \ref{modelpars} and \ref{spotpars}, respectively.
     
\begin{table}[!h]
\caption{Physical parameters of VW Cep according to our \texttt{PHOEBE} light- and radial velocity curve models compared to the previous results of \citet{kaszas98} ($T_\mathrm{eff,1}$, signed with asterisk, was a fixed parameter).}
\label{modelpars}
\centering
\begin{tabular}{c c c} 
\hline\hline
Parameter & This paper & Kaszás et al. (1998) \\
\hline
$q$ & 0.302 $\pm$ 0.007 & 0.35 $\pm$ 0.01\\
$V_{\gamma}$ [km s$^{-1}$] & -12.61 $\pm$ 1.06 & -16.4 $\pm$ 1\\
$a$ [$10^{6}$ km] & 1.412 $\pm$ 0.01 & 1.388 $\pm$ 0.01\\
$i$ [\degr] & 62.86 $\pm$ 0.04 & 65.6 $\pm$ 0.3\\
$T_\mathrm{eff,1}$* [K] & 5050 & 5050\\
$T_\mathrm{eff,2}$ [K] & 5342 $\pm$ 15 & 5444 $\pm$ 25\\
$\Omega_{1}=\Omega_{2}$ & 2.58272 & -- \\
$m_{1}$ [M$_{\odot}$] & 1.13 & 1.01\\
$m_{2}$ [M$_{\odot}$] & 0.34 & 0.36\\
$R_{1}$ [R$_{\odot}$] & 0.99 & --\\
$R_{2}$ [R$_{\odot}$] & 0.57 & --\\
\hline
\end{tabular}
\end{table}   

\begin{table*}
\caption{Spot parameters from the light curve models of single nights.}
\centering
\begin{tabular}{c c c c c c}
\hline \hline
Parameter & 8th August 2014 & 9th August 2014 & 10th August 2014 & 20th April 2016 & 21st April 2016\\ 
\hline
Latitude$_{1}$ [\degr] & 46.8 $\pm$ 4.0 & 48.1 $\pm$ 3.5 & 46.7 $\pm$ 3.9 & 46.1 $\pm$ 2.6 & 45.2 $\pm$ 1.4\\
Longitude$_{1}$ [\degr] & 187.3 $\pm$ 0.02 & 172.2 $\pm$ 0.02 & 179.9 $\pm$ 0.01 & 178.0 $\pm$ 0.01 & 172.8 $\pm$ 0.01\\
Radius$_{1}$ [\degr] & 16.0 $\pm$ 1.0 & 18.1 $\pm$ 0.9 & 17.7 $\pm$ 1.0 & 22.4 $\pm$ 0.6 & 23.0 $\pm$ 0.4\\
Latitude$_{2}$ [\degr] & 46.1 $\pm$ 3.0 & 44.2 $\pm$ 4.8 & 45.4 $\pm$ 1.5 & 45.6 $\pm$ 6.1 & 47.0 $\pm$ 2.5\\
Longitude$_{2}$ [\degr] & 342.3 $\pm$ 0.02 & 352.0 $\pm$ 0.02 & 344.5 $\pm$ 0.01 & 343.4 $\pm$ 0.05 & 338.9 $\pm$ 0.03\\
Radius$_{2}$ [\degr] & 19.5 $\pm$ 0.4 & 20.6 $\pm$ 2.0 & 20.4 $\pm$ 0.5 & 21.0 $\pm$ 0.7 & 20.9 $\pm$ 0.7\\
\hline
\end{tabular}
\label{spotpars}
\end{table*}

\subsection{Spectrum synthesis}
In order to analyze the chromospheric activity of the system, we constructed synthetic spectra for every observed spectra. For spectrum synthesis, we used Robert Kurucz's \texttt{ATLAS9} \citep{kurucz93} model atmospheres selecting different temperatures (4500 $-$ 6500 K, with 250 K steps). The model spectra were Doppler-shifted and convolved with the broadening functions regarding to the observed orbital phases that were computed with the \texttt{WUMA4} program. As in \citet{kaszas98}, the tertiary component is also well-resolved in our cross-correlation functions (Fig. \ref{ccf}), which means that its contribution is not negligible. Accounting for this third component, a $T$ = 5000 K, log $g$ = 4.0 model was used: it was Doppler-shifted, then convolved with a simple Gaussian function applying the FWHM of the transmission function of the spectrograph. A scaled version of this model spectrum \citep[assuming that the third component gives the 10\% of the total luminosity, based on][]{kaszas98} was added to the contact binary model spectrum to mimic the presence of the tertiary component.
We fitted the produced synthetic spectra having different temperatures to the observed data (leaving out the H$\alpha$-region), and found that $T$ = 5750 K and log $g$ = 4.0 gives the best-fit model. This temperature value corresponds to the flux-averaged mean surface temperature, and it is very similar to that found by \citet{hendry00} for the unspotted mean surface temperature of VW Cep. The best-fit model spectra were then subtracted from the observed ones at every epoch. Finally, we measured the equivalent widths of the H$\alpha$-profile on the residual (observed minus model) spectra using the \texttt{IRAF/splot} task. It is hard to fit the H$\alpha$-line of the components with the usual Gaussian- or Voigt-profiles on the residual spectra, so we chose the direct integration of the equivalent widths. We assumed that the uncertainties of the equivalent width measurements mostly come from the error of the continuum normalization, hence they were  determined by the RMS of the relative differences between the observed and synthetic spectra outside the H$\alpha$-region.

\section{The activity of VW Cep}

\begin{figure}
   \centering
   \includegraphics[width=9cm]{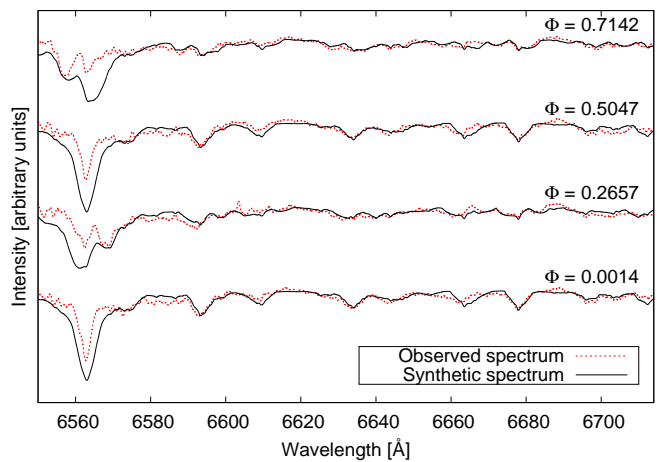}
   \caption{A sample of observed (dashed line) and synthesized (solid line) spectra of VW Cephei in different orbital phases.}
   \label{spectra}
   \end{figure}

A sample of our observed and synthetic spectra is plotted in Fig. \ref{spectra}, while a more detailed view of the H$\alpha$-region along with the difference spectra is available in Appendix D (Fig. \ref{halfa} and \ref{diff}). We can see an obvious excess in the H$\alpha$ line profile, which strongly suggests ongoing chromospheric activity. The excess belongs mainly to the line component of the more massive primary star, while there is no sign of clear emission in the line component of the secondary star. It implies that the chromosphere of the primary star is way more active, and this is consistent with our light curve models where we assumed that spots are located on the primary.

\begin{figure}
   \centering
   \includegraphics[width=9cm]{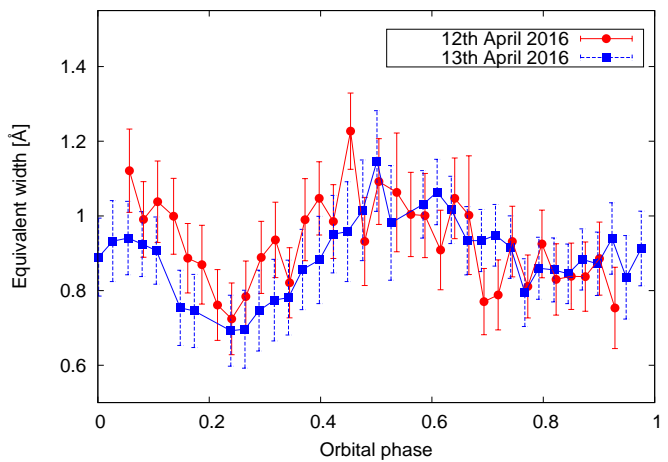}
   \caption{The equivalent widths of the H$\alpha$ emission line profile measured on different nights in the function of orbital phase. For determining the EW values, we subtracted synthetic spectra from the observed ones at every epoch (see the text for details).}
   \label{ew-phase}
   \end{figure}

In Fig. \ref{ew-phase}, we plotted the determined equivalent width values as a function of the orbital phase. As it can be seen, the values on the two consecutive nights in 2016 are almost the same; the small differences may be caused by continuum normalization effects \citep[as noted by][3\% error in the continuum can result in a 0.3 \mbox{\AA} error in the equivalent widths]{kaszas98}.

   \begin{figure}
   \centering
   \includegraphics[width=9cm]{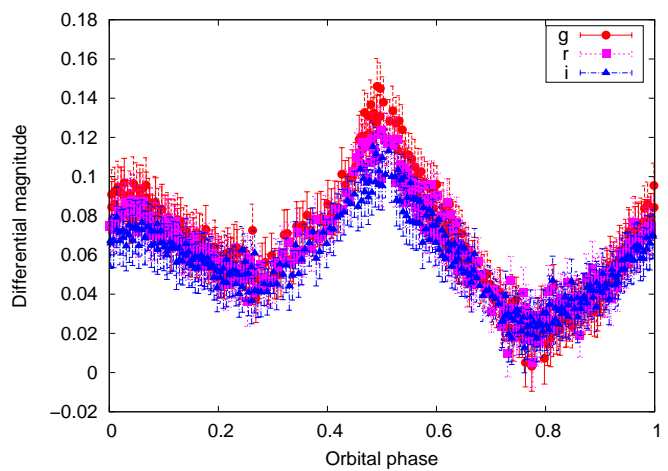}
   \caption{Residuals of light curves measured in 2016 compared to spotless synthetic light curves.}
   \label{res2}
   \end{figure}

The light curves obtained in 2014 (Figs. \ref{20140808lc}, \ref{20140809lc}, \ref{20140810lc}) and 2016 (Figs. \ref{20160420lc}, \ref{20160421lc}) clearly show a smaller maximum brightness at phase $\Phi$ = 0.25. Without spectroscopy, one would most probably think that it is caused by a spot (or a group of spots) on the surface of the primary component when it is moving towards us, or on the secondary component when it is moving away from us. Nevertheless, if we analyze the EWs vs. orbital phase diagram (Fig. \ref{ew-phase}), we can see two peaks centered at the phases of $\Phi$ = 0.1 and $\Phi$ = 0.5. This implies that enhanced chromospheric activity is occurring at these two phases, so we should expect spots there. Hence, we modeled the light curves with two spots (on the regions of the primary that faces the observer at $\Phi$ = 0.1 and $\Phi$ = 0.5 phase, respectively). Better fit was obtained using this two-spots configuration in all three
filter bands than in the case of a single spotted region placed at $\Phi$=0.25 (\texttt{PHOEBE}'s built-in cost function gives an approximately two times higher value  if we use one spot instead of two, keeping the same orbital configuration). Note that the single-spot fit can be improved with increasing the temperature difference of the two stars up to 600 K; however, such high value of $\Delta$T is unlikely in the case of a W UMa star, and the result of the fitting is still worse than assuming two spots. According to the best-fit \texttt{PHOEBE} models, there is a larger spot at $\Phi$ = 0.5 phase when we see a higher peak on the EW-phase diagram (Fig. \ref{ew-phase}), and a smaller spot at $\Phi$ = 0.1 when we see a lower peak. This clearly shows that there is a correlation between spottedness and the strength of the chromospheric activity. In order to visualize this correlation, we subtracted spotless synthetic light curves from the observed light curves obtained in 2016 and plotted the residuals in Fig. \ref{res2}. One can see that the shape of Fig. \ref{res2} is very similar to Fig. \ref{ew-phase}. 

As one can see in Figs. \ref{20140808lc}, \ref{20140809lc}, \ref{20140810lc}, \ref{20160420lc}, and \ref{20160421lc}, the model spots we used are located at relatively high latitudes on the primary component. Although, in general, latitudinal information of spots cannot be reliably extracted from light curve modeling \citep{lanza99}, earlier studies concluded that high-latitude spots are expected on rapidly rotating stars \citep{schussler92, schussler96, holzwarth03}.
In our \texttt{PHOEBE} models spots are situated mainly at two longitudes separated by nearly 180 degrees in line with the center of the components, which is consistent with the model of \citet{holzwarth03}. They showed that this is expected in active close binaries, because the tidal forces can alter the surface distribution of erupting flux tubes and create spot clusters on the opposite sides of the active component. Such active longitudes were found in other close binaries by \citet{olah06}.

This kind of longitudinal distribution of spots could also be the sign of the so-called flip-flop phenomenon, which is explained as the combination of a non-axisymmetric dynamo mode with an oscillating axisymmetric mode in case of a star with spherical shape \citep{elstner05}. This phenomenon has been observed on numerous stars (mainly on FK Com and RS CVn systems, but also on the Sun and on young solar-like stars) that show two active longitudes on their opposite hemispheres with alternating spot activity level \citep[e.g.][]{jetsu93,vida10}. In these cases, the observed variations of the surface activity are somewhat periodic; flip cycles are usually in the range of a few years to a decade. 
Activity cycles of contact binaries have been also assumed previously, based on the cyclic variations of minima seen on O$-$C diagrams, and on the seasonal variations of the light curves, especially of the differences of light curve maxima within a cycle \citep{kaszas98, borkovits05}. Recently, \citet{jetsu16} presented a general model for the light curves of chromospherically active binary stars. According to them, probably all such systems have these two active longitudes with cyclic alternation in their activity (note however, that they analyzed only RS CVn systems). Our data covers a relatively short time and we can't see any large changes in the spot strengths. The only convincing evidence we found is shown in Fig. 1 of \citet{kaszas98}, where the authors plotted the seasonal light curves of VW Cephei. One can see the relative changes in the photometric maxima on a scale of 3-4 years which could be explained by changes in the spot strengths. The only contact binary system with a detected flip-flop phenomenon is HH UMa \citep[based on the seasonal variations of light curves, see][]{wang15}. Nevertheless, the observed light curve variations of HH UMa are very similar to those can be seen in the case of VW Cep \citep[see in Fig. 1 of][]{kaszas98}. Thus, it can be an argument that the presence of the flip-flop phenomenon is a viable hypothesis even in the cases of close binary stars.

\section{O$-$C analysis}

In order to analyze the periodic variations of VW Cep, we constructed the O$-$C diagram of the system. For this purpose, we downloaded all the minima times from the O$-$C gateway webpage\footnote{O$-$C gateway: http://var2.astro.cz/ocgate/} created by Anton Paschke and BC. Lubos Brát. Furthermore, we collected all other available data points from IBVS, BBSAG, BAA-VSS and BAV issues, and, finally, we added the latest points from our measurements (see Table \ref{minima}).

\begin{table}[!h]
\caption{Times of primary and secondary minima obtained from our photometric data.}           
\label{minima}
\centering
\begin{tabular}{c c c} 
\hline\hline
Date of observation & Type of minimum & MJD$_{min}$\\
\hline
08 August 2014 & Primary & 56878.3935\\
08 August 2014 & Secondary & 56878.5342\\
09 August 2014 & Primary & 56879.5061\\
09 August 2014 & Secondary & 56879.3685\\
10 August 2014 & Primary & 56880.3408\\
10 August 2014 & Secondary & 56880.4821\\
20 April 2016 & Primary & 57499.5820\\
20 April 2016 & Secondary & 57499.4451\\
21 April 2016 & Primary & 57500.4172\\
21 April 2016 & Secondary & 57500.5582\\
\hline
\end{tabular}
\end{table}   

\begin{figure}
   \centering
   \includegraphics[width=9cm]{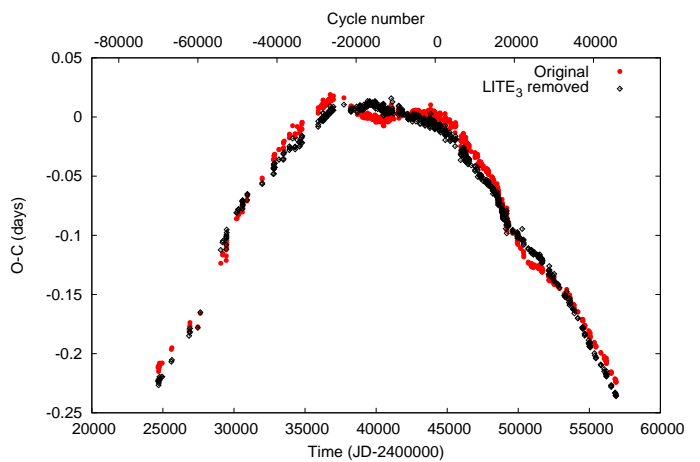}
   \caption{The O$-$C diagram before (red diamonds) and after the subtraction of the visually observed LITE of the 3rd body. Note that cycle numbers are shown on the top horizontal axis.}
   \label{O-C}
   \end{figure}

We only selected the most reliable measurements obtained with photoelectric and CCD devices. For the calculations, we used the reference epoch and period from the GCVS catalogue \citep{kreiner81}:
$${\rm HJD_{min}} = 2444157.4131 + 0.^{d}27831460 \times E.$$
The final diagram, which includes the times of both primary and secondary minima, contains 1620 data points (Fig. \ref{O-C}).

The O$-$C diagram has a clearly visible negative parabolic trend shape. \citet{kaszas98} fitted the long-term period decrease with a parabolic function getting $\Delta P/P$ = $-0.58\times10^{-9}$. They assumed that this phenomenon is caused by a conservative mass transfer from the primary (more massive) component to the secondary one and determined a mass transfer rate of $\Delta M$ = $1.4\times10^{-7}$ M$_{\odot}$ yr$^{-1}$. They conclude that the large rate of period decrease and of the mass transfer indicate that this phenomenon should be temporary.
After the subtraction of the parabola, the residual contains a periodic variation which was explained with the presence of a light-time effect (LITE) by several authors \citep{hill89, kaszas98, pribulla00, zasche07}. This LITE is probably caused by the 3rd body orbiting around the center of mass of the VW Cep AB. \citet{hershey75} confirmed the existence of this tertiary component, but the observed amplitude of the LITE is larger than expected from its orbital elements. After subtracting the effect of the third body, the residual shows additional periodic variations. \citet{kaszas98} proposed two possible explanations for this cyclic behaviour: ($i$) perturbation of orbital elements induced by the tidal force of the the third body, and ($ii$) geometric distortions caused by magnetic cycles \citep[Applegate mechanism;][]{applegate92}.

Regarding the significantly larger number of data points (collected in the last 20 years), we decided to re-analyze the rate of period decrease. Furthermore, we propose new explanations to the different cyclic variations. For these purposes, we removed the light-time effect of the visually observed tertiary component as was also done in \citet[][]{kaszas98}, see Fig. \ref{O-C}.

   \begin{figure}
   \centering
   \includegraphics[width=9cm]{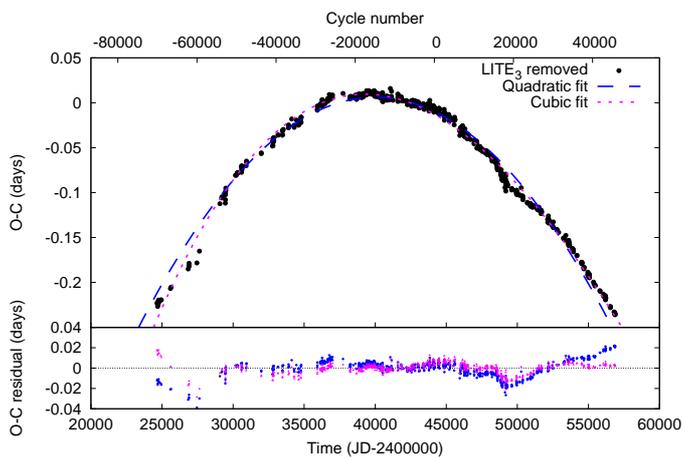}
   \caption{The fitted quadratic (dashed blue curve) and cubic (dotted magenta curve) polynomial to the O$-$C diagram (top) and their residuals (bottom). Note that cycle numbers are shown on the top horizontal axis.}
   \label{O-C poli-cub-fit}
   \end{figure}

As a first approach, the O$-$C diagram was fitted by a quadratic polynomial (Fig. \ref{O-C poli-cub-fit}). The residual shows periodic variations, and, moreover, large deviations in the early years of observations (note that it's difficult to estimate the uncertainties of these values), as well as near and after 20\,000 cycles. Then we fitted the data with a cubic polynomial, which resulted in a smaller sum of squared residuals. The deviations decrease in the intervals mentioned before, which may suggest that the rate of period decrease is changing in time. If we assume that the period decrease is owing to mass transfer, then the period changes and the mass transfer rates can be calculated (see Table \ref{table_fittedpars}), which are in the same order as calculated by \citet{kaszas98}. The cubic fit yielded positive numbers both in the variation of period change and of mass transfer rate, which implies that the efficiency of the mass transfer from the more massive component is decreasing, in agreement with the previous expectations.

\begin{table*}
\caption{O$-$C diagram: fitted (top four rows) and calculated (bottom four rows) parameters. The $a_0$, $a_1$, $a_2$ and $a_3$ are the fitted constant, first-order, second-order and third-order coefficients, respectively. The residuals are the sum of squared residuals. $\dot{P}$ and $\dot{M}$ are the period variation and the mass transfer rate, respectively, while $\frac{d}{dt}\dot{P}$ and $\frac{d}{dt}\dot{M}$ are the temporal variation of the previous parameters, respectively.}
\centering
\begin{tabular}{c c c c}
\hline \hline
& & Quadratic fit & Cubic fit\\ \hline
$a_0$ & - & (-7.02 $\pm$ 0.50) $\times$ 10$^{-3}$ & (-9.64 $\pm$ 0.48) $\times$ 10$^{-3}$\\
$a_1$ & - & (-2.08 $\pm$ 0.02) $\times$ 10$^{-6}$ & (-2.38 $\pm$ 0.03) $\times$ 10$^{-6}$\\
$a_2$ & - & (-7.03 $\pm$ 0.05) $\times$ 10$^{-11}$ & (-6.43 $\pm$ 0.06) $\times$ 10$^{-11}$\\
$a_3$ & - & - & (2.14 $\pm$ 0.14) $\times$ 10$^{-16}$\\
residual & - & 1.30 $\times$ 10$^{-1}$ & 1.03 $\times$ 10$^{-1}$\\
$\dot{P}$ & day yr$^{-1}$ & (-1.85 $\pm$ 0.01) $\times$ 10$^{-7}$ & (-1.69 $\pm$ 0.02) $\times$ 10$^{-7}$\\
$\dot{M}$ & M$_{\odot}$ yr$^{-1}$ & (-1.21 $\pm$ 0.05) $\times$ 10$^{-7}$ & (-1.11 $\pm$ 0.05) $\times$ 10$^{-7}$\\
$\frac{d}{dt}\dot{P}$ & day yr$^{-2}$ &-& (3.0 $\pm$ 0.2) $\times$ 10$^{-12}$\\
$\frac{d}{dt}\dot{M}$ & M$_{\odot}$ yr$^{-2}$ &-& (2.0 $\pm$ 0.2) $\times$ 10$^{-12}$\\
\hline
\end{tabular}

\label{table_fittedpars}
\end{table*}

\begin{figure}
   \centering
   \includegraphics[width=9cm]{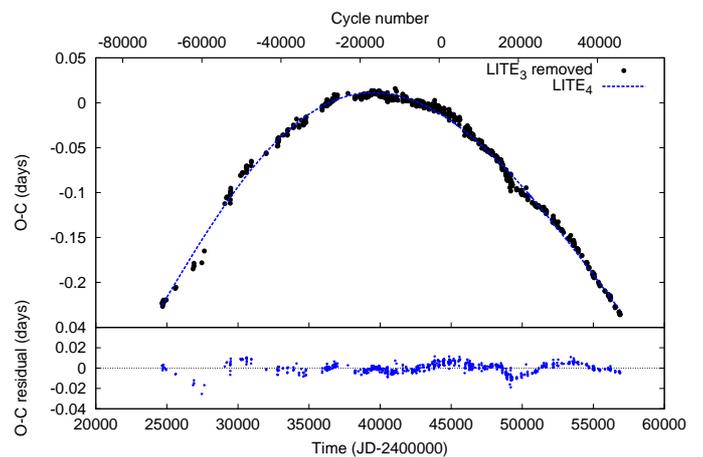}
   \caption{The fitted LITE of the hypothetical 4rd body to the O$-$C diagram without the LITE of the known 3rd body (top) and the residual (bottom). Note that cycle numbers are shown on the top horizontal axis.}
   \label{OCLITE4}
   \end{figure}

The fact that the higher-order polynomial fits better the O$-$C diagram can be the consequence of a periodic variation, which is not covered yet due to its very long time-scale. To utilize this assumption, we subtracted the LITE of the known 3rd body, then we assumed that the long-term variation is caused by a hypothetical 4th body and fitted the residual diagram with a LITE. The best-fit curve generated by using the non-linear least-squares method can be seen in Fig. \ref{OCLITE4}. The fitted parameters are listed in Tab. \ref{tab_Fourier}.

\begin{table}
\caption{Fitted parameters of the LITE of the hypothetical fourth body.}
\begin{center}
\begin{tabular}{c c c}
\hline \hline
T$_0$ & MJD & (2.42 $\pm$ 0.12) $\times$ 10$^4$\\
P & yr & 180.5 $\pm$ 2.0\\
$\omega$ & deg & 0.0 $\pm$ 7.5\\
$e$ & - & 0.114 $\pm$ 0.005\\
$a\sin{i}$ & km & (6482.26 $\pm$ 1.1e-14) $\times$ 10$^6$\\
$f$(M$_4$) & M$_\mathrm{sun}$ & 2.50 $\pm$ 0.06\\
A$_\mathrm{O-C}$ & d & 0.24700 $\pm$ 0.00028\\
\hline
\end{tabular}
\label{tab_Fourier}
\end{center}
\end{table}

Defining $m_{4}$ as the mass of the 4th body, the mass function is
\begin{equation} \label{fm4}
    f(m_{4}) = \frac{m_{4}^{3} \sin{i^{3}}}{(m_\mathrm{VW}+m_{3}+m_{4})^{2}},
\end{equation}
where $m_\mathrm{VW}$ and $m_3$ are the mass of VW Cep AB and of the known 3rd body, respectively, and $i$ is the inclination.
The amplitude of the O$-$C is:
\begin{equation} \label{Aoc}
A_\mathrm{O-C} = \frac{a_{1,2,3} \sin{i}}{c}  \sqrt{(1-e^2 \cos^2{\omega})},
\end{equation}
where $a_{1,2,3}$ is the semi-major axis of the common center of mass of VW Cep AB and the 3rd body around the hypothetic 4th body, $c$ is the speed of light, $e$ is the eccentricity, and $\omega$ is the argument of periapsis.
Combining Eq. \ref{fm4} and Kepler's third law, the mass function of the 4th body can be also calculated as:
\begin{equation}
f(m_4) = 3.99397 \times 10^{-20} \frac{a_{1,2,3}^3 \sin^3{i}}{P_\mathrm{L}^2},
\end{equation}
where the semi-major axis is in km, the period is in days and the mass function yields the result in M$_\odot$.
Calculated semi-major axes and masses of the 4th body (assuming different values of inclination) can be found in Tab. \ref{table_interation}. We used 1.47 M$_{\odot}$ (this paper) and 0.74 M$_{\odot}$ \citep{zasche08} for the total mass of VW Cep AB and for the mass of the 3rd component, respectively.

\begin{table}
\caption{Calculated orbital parameters, masses and proper motions of the hypothetical 4th body assuming different inclination angles.}
\begin{center}
\begin{tabular}{c c c c c c}
\hline \hline
$i$ & $a_{1,2,3}$ & $a_4$ & $m_4$ & $\mu$\\
(\degr) & (10$^6$ km) & (10$^6$ km) & (M$_{\odot}$) & (mas yr$^{-1}$)\\
\hline
10 & 37329.8 & 171.4 & 481.4 & 345.6\\
20 & 18952.9 & 628.5 & 66.6 & 167.6\\
30 & 12964.5 & 1201.0 & 23.9 & 111.5\\
40 & 10084.6 & 1727.3 & 12.9 & 85.9\\
50 & 8462.0 & 2142.9 & 8.7 & 71.7\\
60 & 7485.1 & 2444.3 & 6.8 & 63.3\\
70 & 6898.3 & 2645.7 & 5.8 & 58.3\\
80 & 6582.3 & 2760.8 & 5.3 & 55.6\\
90 & 6482.3 & 2798.1 & 5.1 & 54.7\\
\hline
\end{tabular}
\label{table_interation}
\end{center}
\end{table}

According to celestial mechanics \citep[e.g.][]{reipurth12}, such a four-body system can only be stable if the outer periastron distance is at least 5$-$10 times larger than the inner apastron distance. The semi-major axis of VW~Cep~AB orbit
around the common centre of mass is (647.76 $\pm$ 29.92) $\times$ 10$^6$ km \citep{zasche07}. The semi-major axis of the 3rd body, calculated from the mass ratio m$_3$/m$_{VW}$, is (1199.23 $\pm$ 55.39) $\times$ 10$^6$ km. The sum of the two values is (1846.99 $\pm$ 60) $\times$ 10$^6$ km. If we assume that the distance of the 4th body must be at least 7.5 times \citep[average of the criterion given by][]{reipurth12} larger than the inner system's periastron distance, then we get a$_{1,2,3}$~$\geq$ (13852.4 $\pm$ 500) $\times$ 10$^6$ km. Considering the calculations in Tab. \ref{table_interation}, we can put constrains on inclination, $i$ $\lesssim$ 30\degr, which is in agreement with the assumed inclination of the 3rd body \citep[33.6 $\pm$ 1.2\degr;][]{zasche07}. Using this value, the mass of the 4th body will be at least 23.9 M$_{\odot}$. A star with such large mass should be detectable in the distance of VW Cep, so we note that this object, if exists, should be a black hole. Such a massive object was recently proposed as an explanation of the periodic variation seen on the O$-$C diagram of an RR Lyrae binary candidate in \citet{sodor17}.

The orbital speed of the common center of mass of VW Cep AB and the 3rd body around the center of mass of the four body system is:
\begin{equation}
v = \frac{2\pi}{P_\mathrm{L}}\frac{a_{1,2,3}\sin{i}}{\sqrt{(1-e^2)} \sin{i}}\sqrt{\bigg(1+e^2+2 e \cos{(f)}\bigg)},
\end{equation}
where $f$ is the true anomaly. The projection to the plane of the sky is:
\begin{equation}
v_{sky} = \sqrt{v^2-v_{\mathrm{rad}}^2},
\end{equation}
where
\begin{equation}
v_{\mathrm{rad}} = \frac{2\pi}{P_\mathrm{L}} \frac{a_{1,2,3}\sin{i}}{\sqrt{1-e^2}}\bigg[\cos{(f+\omega)}+e \cos{(\omega)}\bigg],
\end{equation}
the radial velocity, from which the proper motion is:
\begin{equation}
\mu \big[\mathrm{\arcsec\,yr^{-1}}\big] = \frac{v_{\mathrm{sky}} [\mathrm{km\,s^{-1}}]}{4.74 \cdot d [{\rm pc}]}.
\end{equation}

The parallax of the VW Cep AB system is 36.25 $\pm$ 0.58 mas \citep{vanleeuwen07}, which leads to a distance of $27.59_{-0.21}^{+0.48}$ pc. Proper motion values calculated with different inclinations are listed in Tab. \ref{table_interation}. In the literature, there can be found two measured proper motions (Tab. \ref{tab_PM}), which are in the same order as in our calculations assuming a low value of inclination. We checked the proper motion values of the nearby stars in order to investigate if such a large value is common in that field or not. We found only one object (LSPM J2040+7537) with similar order of proper motion. Unfortunately, we haven't found any measured parallax in the literature, but the large angle distance (729.02") and the relatively low brightness (R$=$15.2 mag) suggests that this star is in a large physical distance from VW Cep. Thus, the measured large proper motion values may be partially explained by the presence of a 4th body.

\begin{table}
\caption{Measured proper motion values from literature.}
\centering
\begin{tabular}{c c c}
\hline \hline
$\mu_{\alpha}$ & $\mu_{\delta}$ & Ref.\\
(mas yr$^{-1}$) & (mas yr$^{-1}$) & -\\ \hline
1241.4 $\pm$ 2.7 & 540.9 $\pm$ 0.6 & \citet{vanleeuwen07}\\
1306.5 $\pm$ 16.1 & 551.0 $\pm$ 4.3 & \citet{roeser88}\\
\hline
\end{tabular}
\label{tab_PM}
\end{table}

According to previous papers, the gamma velocity of VW Cephei is continuously changing in time; however, no satisfying explanation have been published for that to date. \citet{zasche08} tried to explain this variation based on the four available gamma velocities in the literature with two different models: ($i$) presence of a third and a fourth component in the system without mass transfer between the primary and secondary, ($ii$) presence of a third component in the system with mass transfer between the primary and secondary. He concluded that the first approach gives a better fit by 12\%, but the fit is still not satisfactory and although the assumed 4th component should be detectable, there are not any candidates yet. More precise data and more sophisticated modeling methods are necessary to solve this problem. However, our new value of $v_{\gamma}$ seems to be in disagreement with the predictions of \citet{zasche08}.

\begin{figure}
   \centering
   \includegraphics[width=9cm]{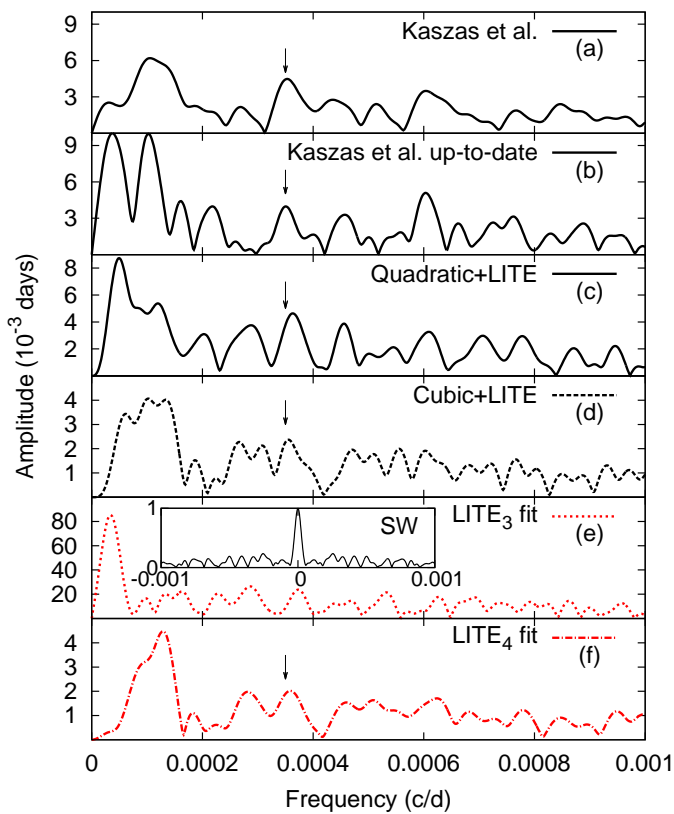}
   \caption{Fourier spectra of the O$-$C diagrams. From top to bottom: (a) after the subtraction of the parabolic trend \citep[following][]{kaszas98} and the visually observed LITE of the 3rd body using data points before MJD 50\,000, (b) same as a), but using the full dataset, (c) after the subtraction of our quadratic fit along with the LITE of the 3rd body, (d) after the subtraction of our cubic fit along with the LITE of the 3rd body, (e) after the subtraction of the LITE of the 3rd body and (f) after the subtraction of the LITE of the hypothetical 4rd body. Note the different ranges of the vertical axes. The insert in the fifth panel shows the spectral window.}
   \label{O-C Fourier}
   \end{figure}

\citet{kaszas98} Fourier-analyzed the residual after subtracting both a parabolic term and a LITE component corresponding to the orbital elements of the 3rd body \citep{hershey75}. They found two significant peaks, one close to the orbital period of the 3rd body ($1.29 \times 10^{-4}$ day$^{-1}$), and another one ($3.59 \times 10^{-4}$ day$^{-1}$) that was linked to the magnetic activity cycle. We recalculated the Fourier spectrum using all published data points up-to-date, and the same fitting parameters as \citet{kaszas98} in order to check whether the mentioned peaks are still visible or not (Fig. \ref{O-C Fourier}; marked with an arrow). Additionally, we calculated the Fourier spectrum after subtracting our quadratic and cubic terms as well as the effect of the 3rd body. As one can see, the shorter period has nearly the same amplitude and frequency in all spectra except the cubic case, where the amplitude is smaller with a factor of 2. The peak in the Fourier spectrum of the data after the subtraction of LITE$_3$ and LITE$_4$ is present with nearly the same parameters as in the case of fitting a cubic function. Regarding the longer period, the amplitude of the corresponding peak is larger; moreover, another peak appears with a period of 73.52 yr in the longer dataset (Fig. \ref{O-C Fourier}. b). Such period caused by the LITE of another hypothetical fourth body has been also proposed by \citet{zasche08}, but in the case of our new quadratic fit, the peak is disappeared or delayed, which means that this can be an artifact.

\citet{applegate92} proposed a mechanism to explain the orbital period changes of eclipsing binaries. Due to the gravitational coupling, the variation of the angular momentum distribution can modify the orbital parameters (such as the semi-major axis, or the orbital period). The shape of the active component of the binary can be periodically distorted by its magnetic cycle of a field of several kilogauss. Such a process cannot be excluded in case of VW Cep because of the photospheric and chromospheric activity of the components observed in the X-ray range \citep{gondoin04, huenemoerder06, sanzforcada07}. We calculated the value of the orbital period modulation ($\Delta P$), the angular momentum transfer ($\Delta J$) and the subsurface magnetic field strength ($B$) required to establish the observed period changes in the O$-$C diagram (Tab. \ref{tab_applegate}) as follows \citep{applegate92}: 

\begin{equation}
\frac{\Delta P}{P} = 2 \pi \frac{A_\mathrm{O-C}}{P_\mathrm{mod}}
\end{equation}
\begin{equation}
\Delta J = \frac{GM^2}{R} \Bigg(\frac{a}{R}\Bigg)^2 \frac{\Delta P}{6 \pi}
\end{equation}
\begin{equation}
B^2 \sim 10 \frac{GM^2}{R^4} \Bigg(\frac{a}{R} \Bigg)^2 \frac{\Delta P}{P_\mathrm{mod}}.
\end{equation}

We used the values of $a$ = 1.412 $\times$ 10$^6$ km (semi-major axis), $P$ = 0.2783 d (orbital period), $M_{1}$ = 1.13 M$_{\odot}$, $M_{2}$ = 0.34 M$_{\odot}$ (masses of the primary and the secondary components), $R_{1}$ = 0.99 R$_{\odot}$, $R_{2}$ = 0.57 R$_{\odot}$ (radii of the primary and secondary components), respectively. The A$_\mathrm{O-C}$ values were calculated using Eq. \ref{Aoc}. and taken from the Fourier-spectrum, respectively. As it can be seen, the calculated subsurface magnetic field strength is lower in case of the shorter modulation. In Section 3, we showed that the primary, more massive component is covered by cool spots, emerged as results of the magnetic activity. If we assume that the 7.62 year long modulation in the Fourier spectrum is linked to the magnetic cycle of VW Cep, then the required subsurface magnetic field strength for establishing the variation is higher than 20 kG. If this field strength becomes lower even by 2-3 orders of magnitude at the surface, it should be detectable with spectropolarimetric observations \citep[see][]{donati09}; however, the detection of strength and variation of this magnetic field (via e.g.  Zeeman effect) would be a challenging task due to the large rotational broadening of spectral lines (approx. 3.3 {\AA} at 5000~\AA).

\begin{table*}
\caption{The calculated angular momentum transfer ($\Delta J$) and the subsurface magnetic field strength ($B$) required to establish the observed period changes.}
\begin{center}
\begin{tabular}{c c c c c c c}
\hline \hline
$A_\mathrm{O-C}$ & $P_\mathrm{mod}$ & $\Delta P/P$ & $\Delta P$ & $\Delta J$ & $B$ & Component\\
(d) & (yr) & - & (sec/cycle) & (g cm$^2$ s$^{-1}$) & (kG) & -\\ \hline


\multirow{2}{*}{0.247} & \multirow{2}{*}{180} & \multirow{2}{*}{2.36 $\times$ 10$^{-5}$} & \multirow{2}{*}{0.57} & 6.26 $\times$ 10$^{47}$ & 8.0 & 1\\
& & & & 2.97 $\times$ 10$^{47}$ & 12.6 & 2\\ \hline


\multirow{2}{*}{0.004} & \multirow{2}{*}{7.62} & \multirow{2}{*}{9.04 $\times$ 10$^{-6}$} & \multirow{2}{*}{0.22} & 2.39 $\times$ 10$^{47}$ & 24.0 & 1\\
& & & & 1.14 $\times$ 10$^{47}$ & 37.8 & 2\\ \hline
\hline
\end{tabular}
\label{tab_applegate}
\end{center}
\end{table*}

\section{Summary}

We obtained 69 new medium-resolution spectra of VW Cephei along with 2 new light curves in April 2016, and with 3 additional light curves from two years earlier, in order to analyze the period variation and the apparent activity of the system. We modelled our new light curves and radial velocity curve simultaneously using the \texttt{PHOEBE} code and also derived 10 new times of minima, which were used to construct a new, extended O$-$C diagram of the system. We redetermined the physical parameters of the system from our models, which are consistent with those in \citet{kaszas98}. All observed spectra were compared to synthetic spectra, then we measured the equivalent widths of the H$\alpha$ profile on every difference spectra. According to the light curve and spectral models, it is confirmed that both photospheric and chromospheric activity is mostly occuring on the more massive primary component of the system. Based on our \texttt{PHOEBE} models, the spot distribution seems to be stable, and spots are located on the two opponent hemispheres of the primary in line with the center of both components. The spots also have different sizes or different temperatures which causes the O'Connell effect on the light curves. The equivalent widths of the H$\alpha$ line show enhanced chromospheric activity in that two phases where the spots actually turn in our line of sight, which means that there is a correlation between spottedness and the strength of the chromospheric activity. We proposed that this kind of spot distribution might be a consequence of the so-called flip-flop phenomenon, which was recently detected on an other contact binary, HH UMa by \citet{wang15}. We noted that the changes on the light curves of different seasons in Fig. 1 of \citet{kaszas98} can also support this hypothesis. Nevertheless, more observations are still needed for confirmation, especially both photometric and spectroscopic measurements taken simultaneously.

After subtracting the LITE of the previously known 3rd component from the O$-$C diagram, our period analysis showed that the mass transfer rate is decreasing, which means that this effect is expected to be temporary as it was proposed earlier by \citet{kaszas98}. In the Fourier spectrum of the O$-$C diagram, we found a significant peak at about 185 years and also detected a peak at 7.62 years, which was connected to the magnetic activity cycle by \citet{kaszas98}. We tested two previously suggested hypothesis to explain these peaks: ($i$) the presence of an additional fourth component in the system, ($ii$) ongoing geometrical distortions by magnetic cycle. The supporting fact of the former is that a cubic polynomial fits the O$-$C diagram better than a quadratic one. This could indicate a longer periodic variation that is not fully covered yet by the observations. We treated the long-term variation as LITE and assumed that this might be a consequence of the presence of a fourth body, which mass would be at least 23.9 M$_{\odot}$ (assuming $i$ $\lesssim$ 30\degr). This scenario could also be a partial explanation of the large proper motion values of VW Cep. The other scenario we analyzed is that the period variation could be the consequence of the Applegate-mechanism; from our calculations, the necessary strength of the subsurface magnetic field required to maintain the observed period variation is higher than 20 kG. Nevertheless, the available data suggest that mass transfer from the more massive primary to the less massive secondary star (with a slowly decreasing mass transfer rate) is the most probable explanation of the observed period variation of VW Cep.

\section*{Acknowledgments}
We would like to thank our anonymous referee for his/her valuable suggestions, which helped us to improve the paper. This project has been supported by the Lendület grant LP2012-31 of the Hungarian Academy of Sciences and the Hungarian National Research, Development and Innovation Office, NKFIH-OTKA K113117 grant. K. Vida acknowledges the Hungarian National Research, Development and
Innovation Office
grants OTKA K-109276, and supports through
the Lend\"ulet-2012 Program (LP2012-31) of the Hungarian Academy of
Sciences,
and the ESA PECS Contract No. 4000110889/14/NL/NDe.
K. Vida is supported by the Bolyai J\'anos Research Scholarship of the
Hungarian Academy of Sciences.

\begin{appendix}

\section{Light curves and spot configurations}

   \begin{figure}[!ht]
   \centering
   \includegraphics[width=9cm]{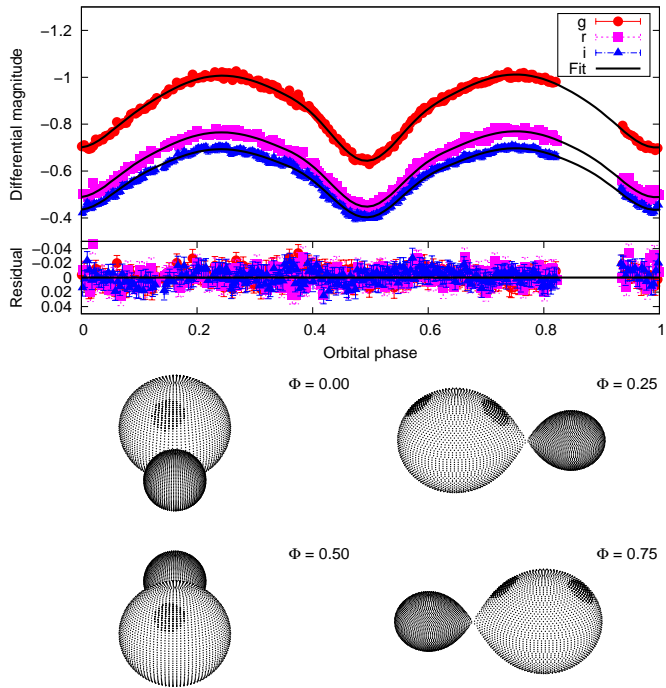}
   \caption{SDSS {\it g'r'i'} light curves of VW Cephei on 9th August 2014 along with the fitted \texttt{PHOEBE} models and the corresponding geometrical configuration and spot distribution in different orbital phases.}
   \label{20140809lc}
   \end{figure}
     
   \begin{figure}[!ht]
   \centering
   \includegraphics[width=9cm]{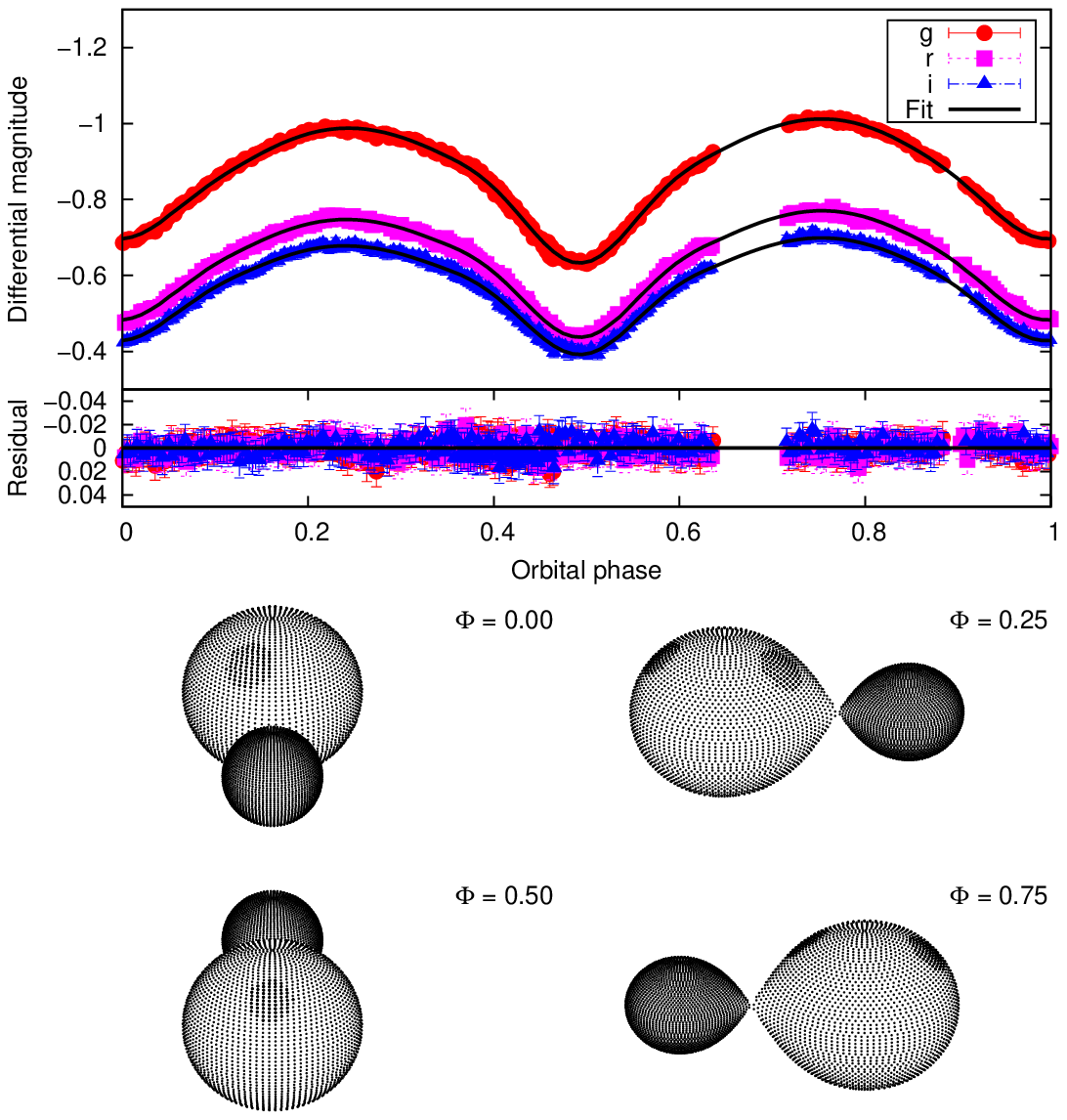}
   \caption{SDSS {\it g'r'i'} light curves of VW Cephei on 10th August 2014 along with the fitted \texttt{PHOEBE} models and the corresponding geometrical configuration and spot distribution in different orbital phases.}
   \label{20140810lc}
   \end{figure}
  
   \begin{figure}[!ht]
   \centering
   \includegraphics[width=9cm]{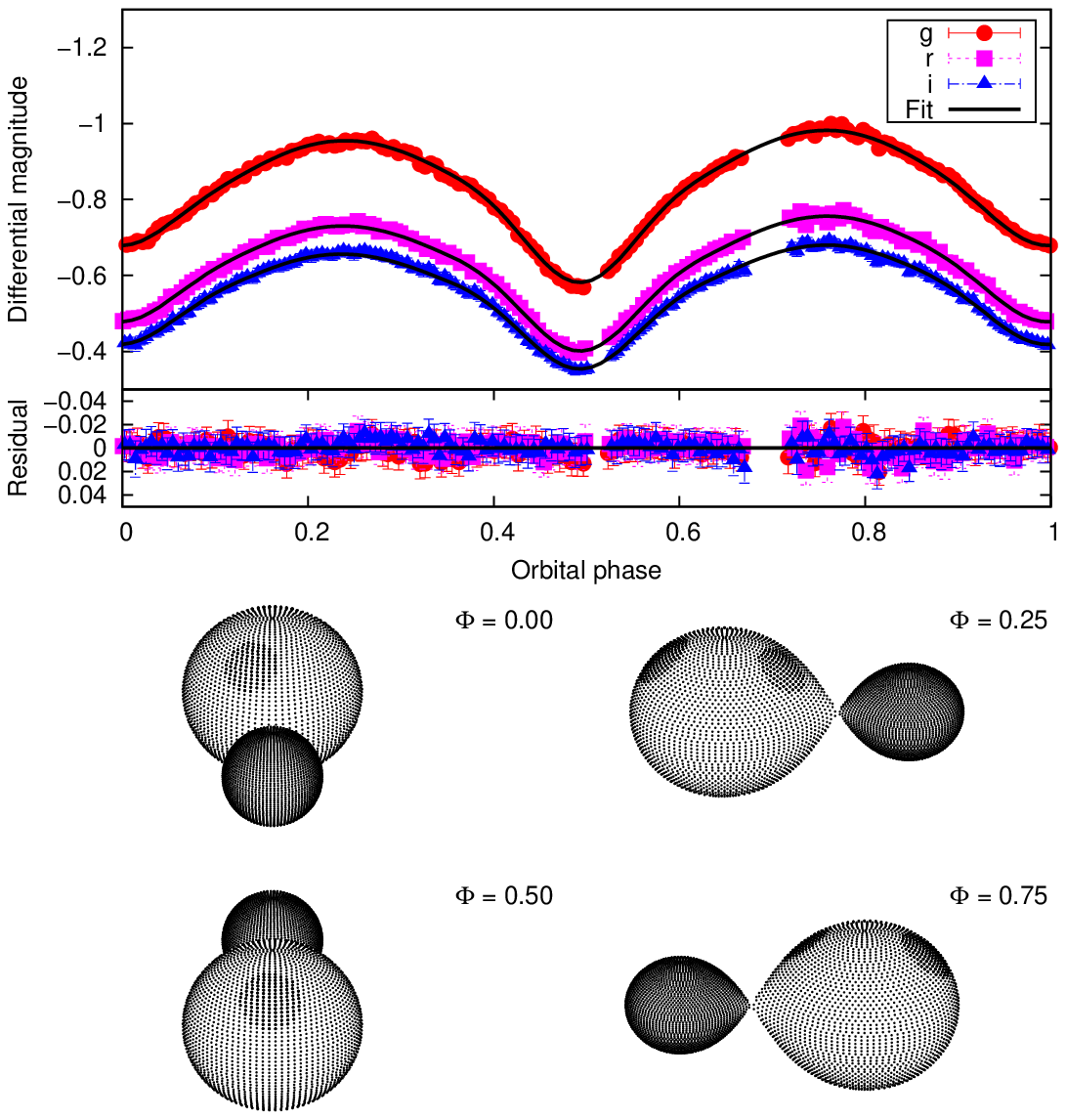}
   \caption{SDSS {\it g'r'i'} light curves of VW Cephei on 20th April 2016 along with the fitted \texttt{PHOEBE} models and the corresponding geometrical configuration and spot distribution in different orbital phases.}
   \label{20160420lc}
   \end{figure}

   \begin{figure}[!ht]
   \centering
   \includegraphics[width=9cm]{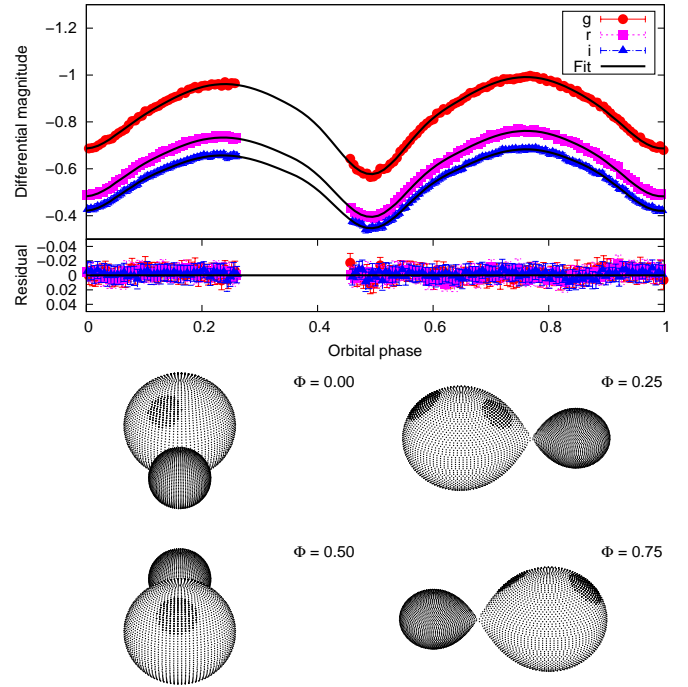}
   \caption{SDSS {\it g'r'i'} light curves of VW Cephei on 21st April 2016 along with the fitted \texttt{PHOEBE} models and the corresponding geometrical configuration and spot distribution in different orbital phases.}
   \label{20160421lc}
   \end{figure}

\newpage

\section{Checking the limitations of the cross-correlation technique}
 
\begin{figure}
   \centering
   \includegraphics[width=9cm]{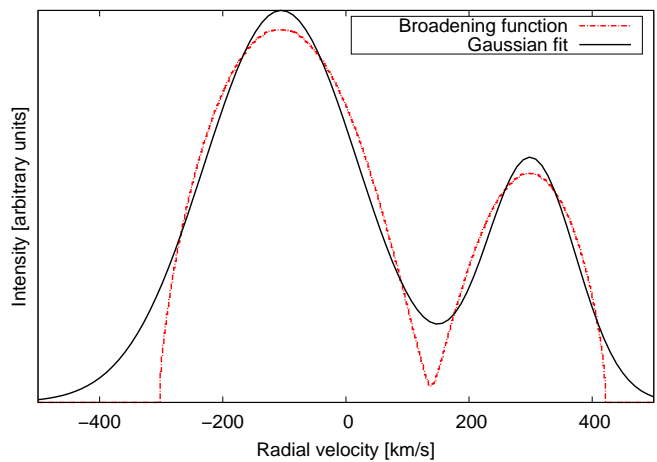}
   \caption{A broadening function of the $\Phi$ = 0.2642 phase and its fit with the sum of two Gaussian functions.}
   \label{bf_fit}
   \end{figure}

\begin{figure}
   \centering
   \includegraphics[width=9cm]{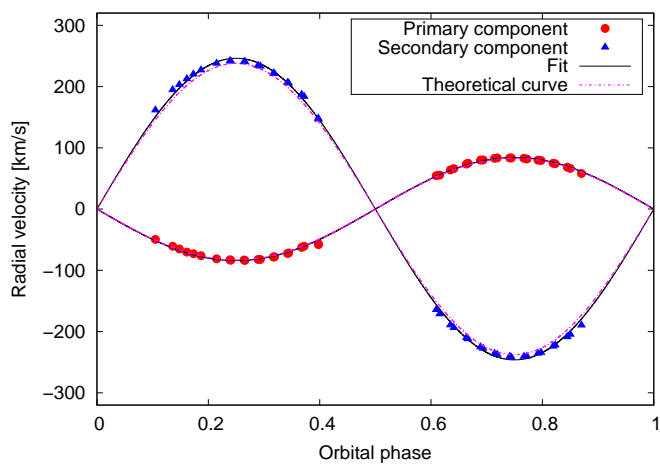}
   \caption{Radial velocity curve of VW Cephei measured on the theoretical broadening functions. The solid line shows the fitted sine functions and the dashed line shows the theoretical velocity curve.}
   \label{bf_radseb}
   \end{figure}

 We used the parameters derived by \citet{kaszas98} for these calculations. A broadening function of the $\Phi$ = 0.2642 phase is plotted along with the fitted Gaussian functions in Fig. \ref{bf_fit}. The radial velocity points measured at the maxima of the two components of the broadening function are plotted in Fig. \ref{bf_radseb} along with the best-fit sine curves (solid line) and the theoretical curve of the system computed from the input parameters (dashed line). There are only minor differences between the two curves, and the measured mass ratio ($q_\mathrm{measured}$ = 0.344) has only a 1.8 \% relative difference compared to the original value ($q_\mathrm{original}$ = 0.35).

\section{Radial velocities}

\begin{table}[!ht]
\caption{The list of newly derived radial velocities and their errors.}
\begin{center}
\begin{tabular}{c c c c}
\hline \hline
Phase & $v_{1}$ [km s$^{-1}$] & $v_{2}$ [km s$^{-1}$] & S/N\\
\hline
0.1361 & -72.99 $\pm$ 1.80 & 182.18 $\pm$ 2.04 & 34.68 \\
0.1479 & -80.43 $\pm$ 1.82 & 195.87 $\pm$ 2.21 & 32.03 \\
0.1615 & -76.45 $\pm$ 1.69 & 196.82 $\pm$ 1.95 & 34.76 \\
0.1733 & -73.79 $\pm$ 1.82 & 208.95 $\pm$ 2.33 & 33.08 \\
0.1869 & -86.67 $\pm$ 2.42 & 213.07 $\pm$ 2.61 & 36.15 \\
0.2149 & -89.12 $\pm$ 2.35 & 228.91 $\pm$ 2.38 & 34.46 \\
0.2388 & -85.55 $\pm$ 1.97 & 239.79 $\pm$ 2.25 & 32.79 \\
0.2403 & -90.31 $\pm$ 2.35 & 237.76 $\pm$ 2.43 & 31.53 \\
0.2642 & -83.77 $\pm$ 1.82 & 236.11 $\pm$ 2.17 & 33.67 \\
0.2657 & -90.20 $\pm$ 2.01 & 235.59 $\pm$ 2.30 & 35.28 \\
0.2895 & -82.89 $\pm$ 1.80 & 232.39 $\pm$ 2.19 & 34.03 \\
0.2934 & -83.48 $\pm$ 1.91 & 231.55 $\pm$ 2.41 & 31.39 \\
0.3170 & -86.73 $\pm$ 2.28 & 221.66 $\pm$ 2.70 & 33.18 \\
0.3188 & -81.57 $\pm$ 2.11 & 225.88 $\pm$ 2.27 & 29.98 \\
0.3424 & -74.94 $\pm$ 1.83 & 212.17 $\pm$ 2.34 & 34.29 \\
0.3442 & -76.54 $\pm$ 1.90 & 210.43 $\pm$ 2.27 & 34.42 \\
0.3677 & -68.99 $\pm$ 2.18 & 194.99 $\pm$ 3.02 & 33.65 \\
0.3721 & -69.15 $\pm$ 1.61 & 193.59 $\pm$ 2.18 & 34.79 \\
0.6094 & 9.90 $\pm$ 1.48 & -207.28 $\pm$ 2.10 & 32.79 \\
0.6155 & 22.91 $\pm$ 2.23 & -203.36 $\pm$ 3.08 & 35.58 \\
0.6348 & 30.43 $\pm$ 1.22 & -212.34 $\pm$ 1.55 & 33.11 \\
0.6409 & 37.24 $\pm$ 1.53 & -215.81 $\pm$ 1.87 & 35.21 \\
0.6634 & 47.40 $\pm$ 1.46 & -225.34 $\pm$ 1.74 & 33.61 \\
0.6663 & 46.40 $\pm$ 1.67 & -233.08 $\pm$ 1.98 & 34.56 \\
0.6888 & 55.00 $\pm$ 1.52 & -235.05 $\pm$ 1.89 & 33.83 \\
0.6935 & 54.99 $\pm$ 1.60 & -243.48 $\pm$ 1.82 & 34.09 \\
0.7142 & 57.64 $\pm$ 1.62 & -245.91 $\pm$ 1.98 & 34.79 \\
0.7189 & 63.16 $\pm$ 1.70 & -257.26 $\pm$ 3.15 & 27.75 \\
0.7414 & 55.45 $\pm$ 1.60 & -254.58 $\pm$ 2.08 & 34.88 \\
0.7443 & 56.60 $\pm$ 1.83 & -258.58 $\pm$ 2.18 & 35.67 \\
0.7668 & 58.28 $\pm$ 1.72 & -249.35 $\pm$ 2.01 & 34.04 \\
0.7722 & 57.04 $\pm$ 1.92 & -252.18 $\pm$ 2.08 & 37.41 \\
0.7922 & 56.19 $\pm$ 1.58 & -247.86 $\pm$ 1.82 & 34.47 \\
0.7976 & 62.71 $\pm$ 1.76 & -245.60 $\pm$ 1.86 & 36.28 \\
0.8193 & 57.16 $\pm$ 1.92 & -238.92 $\pm$ 2.09 & 34.38 \\
0.8230 & 51.16 $\pm$ 1.76 & -238.85 $\pm$ 1.90 & 36.11 \\
0.8447 & 53.48 $\pm$ 2.26 & -222.43 $\pm$ 2.37 & 33.97 \\
0.8501 & 47.71 $\pm$ 2.10 & -230.49 $\pm$ 2.22 & 35.42 \\
0.8701 & 45.21 $\pm$ 2.17 & -206.96 $\pm$ 2.62 & 33.99 \\
\hline
\end{tabular}
\label{radseb}
\end{center}
\end{table}

\newpage

\section{H$\alpha$ profiles and difference spectra}

   \begin{figure}[!ht]
   \centering
   \includegraphics[width=9cm]{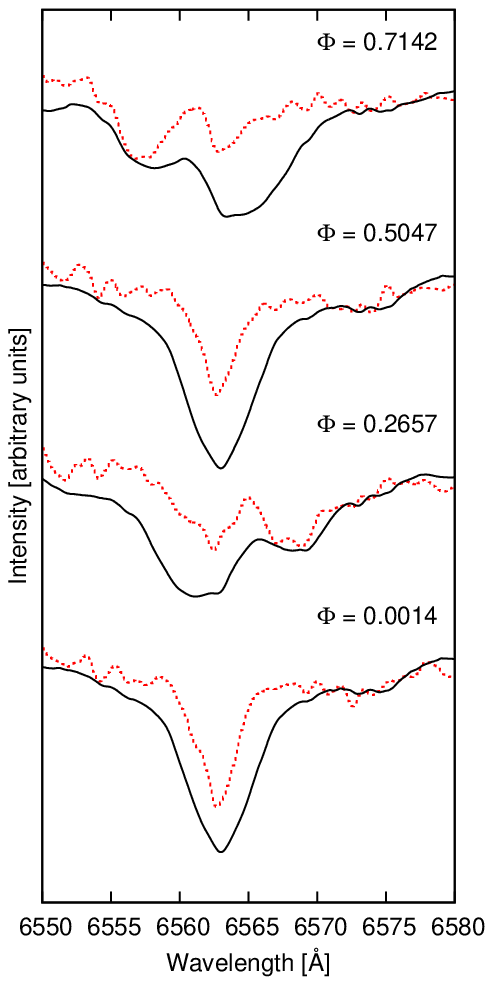}
   \caption{The H$\alpha$ region of the previous sample of observed and synthesized spectra of VW Cephei in different orbital phases.}
   \label{halfa}
   \end{figure}

   \begin{figure}[!ht]
   \centering
   \includegraphics[width=9cm]{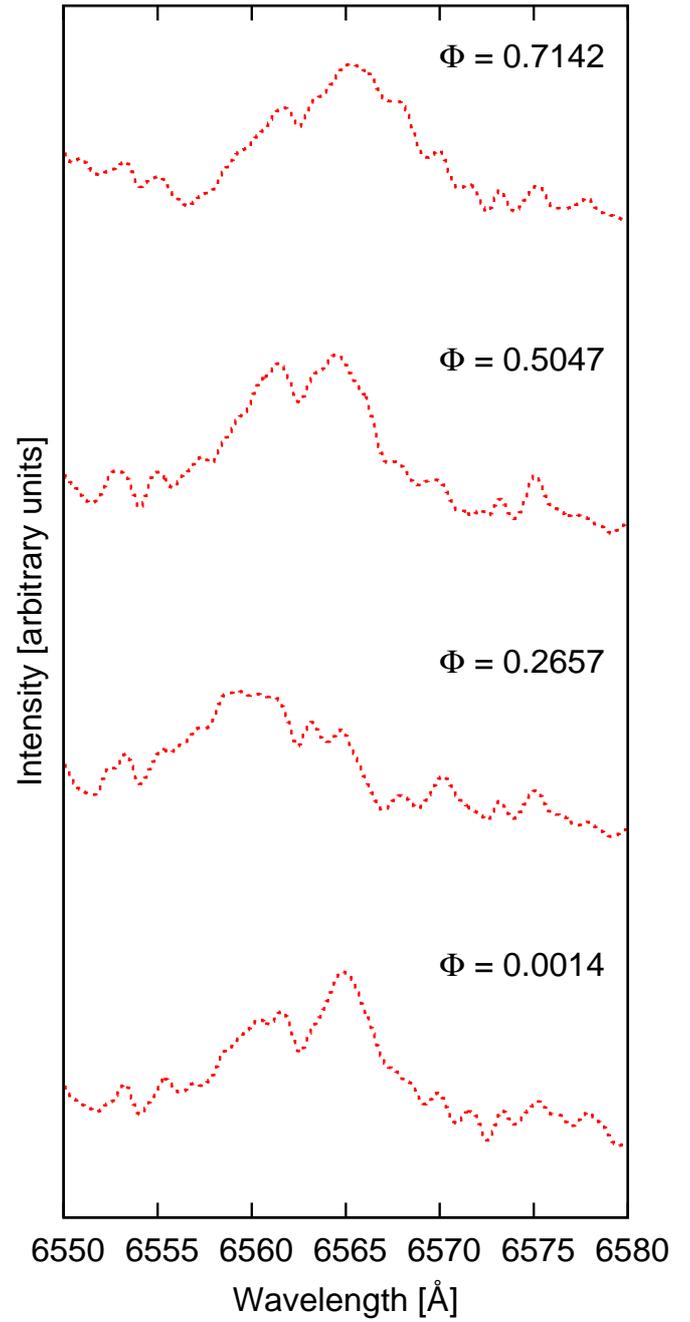}
   \caption{The difference spectra around the H$\alpha$ region of the previous sample of observed and synthesized spectra of VW Cephei in different orbital phases.}
   \label{diff}
   \end{figure}
   
\end{appendix}

\end{document}